# Clinical Relationships Extraction Techniques from Patient Narratives

Wafaa Tawfik Abdel-moneim[1], Mohamed Hashem Abdel-Aziz[2], and Mohamed Monier Hassan[3]

[1]Information System Department, Faculty of Computers and Informatics, Zagazig University

Banha, Egypt

[2]Information System Department, Faculty of Computers and Informatics, Ain-Shames University

Cairo, Egypt

[3]Information System Department, Faculty of Computers and Informatics, Zagazig University

Zagazig, Egypt

**Abstract**

The Clinical E-Science Framework (CLEF) project was used to extract important information from medical texts by building a system for the purpose of clinical research, evidence-based healthcare and genotype-meets-phenotype informatics. The system is divided into two parts, one part concerns with the identification of relationships between clinically important entities in the text. The full parses and domain-specific grammars had been used to apply many approaches to extract the relationship. In the second part of the system, statistical machine learning (ML) approaches are applied to extract relationship. A corpus of oncology narratives that hand annotated with clinical relationships can be used to train and test a system that has been designed and implemented by supervised machine learning (ML) approaches. Many features can be extracted from these texts that are used to build a model by the classifier. Multiple supervised machine learning algorithms can be applied for relationship extraction. Effects of adding the features, changing the size of the corpus, and changing the type of the algorithm on relationship extraction are examined.

## 1. Introduction

Text mining can be defined as a knowledge-intensive process in which user deal with a document collection over time to extract useful and previously unknown information from data sources by using a suite of analysis tools. Text mining deals with the documents that are found in unstructured textual data. Text mining involves the application of techniques from areas such as Information Retrieval (IR), Natural Language Processing (NLP), Information Extraction (IE) and Data Mining (DM). NLP is commonly divided into several layers of processing: lexical, syntactic, and semantic level. The lexical level processing deals with words that can be recognized, analyzed, and identified to enable further processing. The syntactic level analysis deals with identification of structural relationships between groups of words in sentences, and the semantic level is concerned with the content-oriented perspective or the meaning attributed to the various entities identified within the syntactic level [1]. Natural Language Processing (NLP) has been widely applied in biomedicine, particularly to improve access to the ever-burgeoning research literature. Increasingly, biomedical researchers need to relate this literature to phenotypic data: both to populations, and to individual clinical subjects. The computer applications used in biomedical research therefore need to support genotype-meets-phenotype informatics and the move towards translational biology. This will undoubtedly include linkage to the information held in individual medical records: in both its structured and unstructured (textual) portions. Information extraction is the process of automatically obtaining structured data from an unstructured natural language document. Often this involves defining the general form of the information that are important in as one or more templates, which then are used to guide the extraction process. IE systems rely heavily on the data generated by NLP systems [2].

Information extraction system contains information such extract relations between entities from texts. Figure 1 contains example of relation mentions from the news data sets. The left side of the figure contains a pipeline representation of the RE task. The input consists of natural language documents containing e.g. unstructured text or speech. These documents are fed to the RE system, which identifies and characterizes the relations described in the text or speech data. The output of the RE system consists of relation mention triples which include the two entity mentions that take part in the relation and the relation type. The right side of Figure 1 contains example input document on the top and the relation mention triples from these sentence on the bottom. The document contains the sentence "George Bush traveled to France on Thursday for a summit". This contains relation mentions: a reference to a Physical.Located relation between "George Bush" and "France".

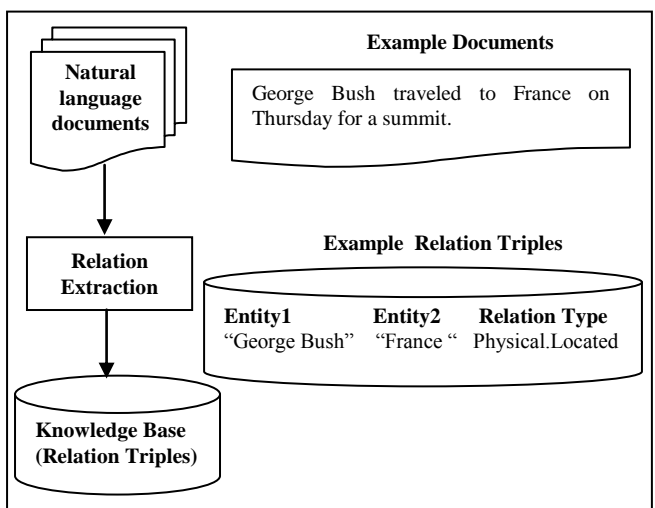

Fig. 1 Overview of relation extraction task with example input and output.

For processing the clinical information, a framework is defined which called the Clinical E-Science Framework (CLEF) project [3] for capture, integration and presentation of this information. The project's data resource is clinical narratives of the cancer patients from The University of Chicago medicine and Mayo Clinic. Information Extraction (IE) technology can be used by CLEF project to extract important information from clinical text. Entities, relationships and modifiers can be extracted from text by the CLEF IE system. The purpose of extracting these relationships is obtaining important information that is often not available in the structured record. What were the interventions for treating the problems? Where was the disease located? Which drugs were given for treating the problems? The extracted relationships are very important for clinical and research applications of IE. The supervised learning algorithm has two properties. The first property is that it does not learn any training example until an unseen example is given; it is called lazy based learning algorithm [4]. The second property is that it classified unseen objects based on target labels of their similar samples; it is called example based learning algorithm [4]. The project uses the guidelines of the CLEF IE system that concerned with extracting entities [5]. Gold standard – human annotated documents – is used to build models of patient narratives which can be applied to unseen patient files. This paper focuses on extracting relationships from patient narratives. Our approach uses different supervised learning algorithms such Support Vector Machine (SVM), Perceptron Algorithm with Uneven Margins (PAUM), NaiveBayes Weka, KNN Weka, and C4.5 Weka classifiers to extract these relationships. The classifiers use a gold standard corpus of oncology narratives which hand annotated with entities and relationships for system training and evaluation. A lot of experiments are applied to reach the much suitable algorithm that gives the high accuracy. The experiments are used different feature sets and see their effects on the system performance. These features sets derived from a linguistic analysis and syntactic analysis of sentence. Relationship is extracted in the same sentence which called inter-sentential relationships. Examine the influence of changing training corpus size for relationship extraction.

## 2. Previous Work

Now a day it is easy to store large amounts of data. Documents are available on the web, intranets, and news wires. However, amount of data available to us are still increasing, our ability is to extract the useful information from this data. Text mining is a good technique to extract useful information from texts. There are many forms of the useful information that can be extracted from the texts such as entities, events, attributes and facts. This information is helpful for researchers to understand the texts very easy. Information extraction (IE) framework has the practical goal of extracting structured information from natural language [6]. IE as a task was formalized largely in the context of the Message Understanding Conference (MUC) shared tasks (e.g., [7]; [8]). Message Understanding Conference (MUCs) [9] describes a method of classifying facts (information) into categories or levels. The researchers develop MUC project by adding new part that contains extracting relationships between entities that take place in MUC-7 [10] such employee_of, product_of, location_of.

In MUC-7, the training examples can be analyzed and hand annotated by the researchers to match contexts which expressed the relevant relation. The work in MUC can be classified into several dimensions: the text type (e.g. newswire, scientific papers, clinical reports); the relations addressed (e.g. part-of, located-in, protein-protein interaction); and the techniques used (e.g. rule-based engineering techniques, supervised learning techniques). In the case of rule-based engineering, writing extraction rules requires extensive effort from a rule engineering expert who is familiar with the target domain. In the case of supervised learning, annotation of training data and features/model parameters require extensive effort from at least one annotator (expert in the target domain) and from a natural language processing expert. In addition to MUC can be classified according to annotated corpora and evaluation software exist (e.g. the ACE relation extraction challenges [11], the LLL genic interaction extraction challenge [12], the BioCreative-II protein-protein interaction task [13]). Many systems use a syntactic parse with domain-specific grammar rules such Linguistic String project [14] to fill template data structures corresponding to medical statements. Other systems use a semantic lexicon and grammar of domain-specific semantic patterns such MedLEE [15] and BioMedLEE [16] to extract the relationships between entities. Other systems use a dependency parse of texts such MEDSYNDIKATE [17] to build model of entities and their relationships. MENELAS [18] also use a full parse. All these approaches are knowledge-engineering approaches. In addition to supervised machine learning has been applied to clinical text. There are many works on relation extraction from biomedical journal papers and abstracts. This work has been done within the hand-written rule base/knowledge engineering approaches.

Now a days the work on relation extraction using supervised ML techniques to train relation classifiers on human annotated texts. The annotated texts contain relation instances which contain relation type and their pair entities. There are many different approaches according to the ML algorithms and the features applied. There are several applications work on biomedicine such using maximum entropy approaches [19], conditional random fields [20] and rule learning methods such as boosted wrapper induction and RAPIER [21] and inductive logic programming [22]. SVMs also have been used for relation extraction [23] but not widely in biomedicine applications. Additional examples for relation extraction contains on [24], [25], and [26]. Currently researchers extract relations from clinical text (such patient narratives) using wide range of features by supervised ML approaches such as SVMs classifiers [27] and [28]. Relationships can be extracted from clinical texts as a part of clinical IE system. Different supervised Machine Learning approaches can be applied to show their affecting on the classification tasks specifically in relation extraction. Several features are used to extract relations such lexical, syntactic, and semantic features.

## 3. Techniques

### 3.1 Relationship schema

Firstly, this application focuses on extracting entities, relationships and modifiers from text. The real thing or event can be found in text documents called entity. Examples of entities in clinical text include diseases, location and drugs and so on. The words that describe an entity called modifiers such as the negation of a condition ("no sign of cancer"), the sub_location of an anatomical locus ("superior-vena-caval "). Relationships are entities that connected to each other and to modifiers e.g. linking an investigation to its result (CT scan shows no abnormality), linking the condition to an anatomical locus (back pain), and linking laterality to an anatomical locus (right groin). Entities, modifiers, and relationships can be determined in XML schema with hand- annotated. Table 1 shows relationship extraction and their argument types from clinical text with a description and examples of each type. Unified Medical Language System (UMLS) semantic network [29] was used to map each entity type into several UMLS types. Relationship shows the clinical dependencies between entities in patient narrative. The schema of relationship was described in [30]. Relationship is linking between pairs of specific entity types, e.g. the Has_finding relation link between investigation and result or between investigation and condition. The schema of relationship types and their argument types is shown graphically in Figure 2.

### 3.2 Gold standard corpus

Entities and relationships in oncology narratives can be hand-annotated by the schema and definitions to provide a gold standard for system training and evaluation. Narratives refer to notes, letters, and summaries written by the oncologist that describe the patient' care. Given the expense of human annotation, the gold standard portion of the corpus has to be a relatively small subset of the whole corpus. In order to avoid events that are either rare or outside of the main project requirements, it is restricted by diagnosis, and only considers documents from those patients with a primary diagnosis code in one of the top level sub-categories of ICD-10 Chapter II (neoplasms) [30]. In order to ensure even training and fair evaluation across the entire corpus, Narratives were selected by randomised and stratified sampling from a larger population of documents. The corpus contains 40 narratives, which were carefully selected and annotated according to a best approach, as described in [30]. This corpus is clinical narratives of the cancer patients from The University of Chicago medicine [31] and Mayo Clinic hospital [32].

Table 1: Description of relationship types and their argument

| Relationship type | First argument type | Second argument type | Description | Example |
|---|---|---|---|---|
| Has_target | Investigation, Intervention | Locus | Relates an intervention or an investigation to the bodily locus at which it is targeted. | • This patient has had a [arg1] bowel [arg2] ultrasound.<br>• This patient has had a [ard2] chest [arg1] X-ray. |
| Has_finding | Investigation | Condition, Result | Relates a condition to an investigation that demonstrated its presence, or a result to the investigation that produced that result. | • This patient has had a [arg1] Ultrasound scanning which shows [arg2] hydronephrosis.<br>• A chest [arg1] X-ray was [arg2] normal. |
| Has_indication | Drug or device, Intervention, Investigation | Condition | Relates a condition to a drug, intervention, or investigation that is targeted at that condition | • … [arg1] chemotherapy to treat the [arg2] cancer.<br>• [arg1] remove brain [arg2] tumors. |
| Has_location | Condition | Locus | Relationship between a condition and a locus: describes the bodily location of a specific condition. | • This patient has had a [arg1] benign cyst on her [arg2] thyroid.<br>• This patient has had a [arg1] lung [arg2] cancer. |
| Modifies | Negation signal | Condition | Relates a condition to its negation or uncertainty about it. | • There was [arg1] no signs of the [arg2] tumor.<br>• There was [arg1] no-evidence of superior-vena-caval [arg2] obstruction. |
| Modifies | Laterality signal | Locus, Intervention | Relates locus or intervention to its sidedness: right, left, bilateral. | •…in her [arg1] left [arg2] breast.<br>• [arg1] bilateral [arg2] mastectomies. |
| Modifies | Sub-location signal | Locus | Relates locus to other information about the location: upper, lower, extra- etc. | • [arg1] lower [arg2] lung.<br>• [arg1] outside the [arg2] prostate.<br>• This patient suffers from [arg1] upper [arg2] abdominal pain. |

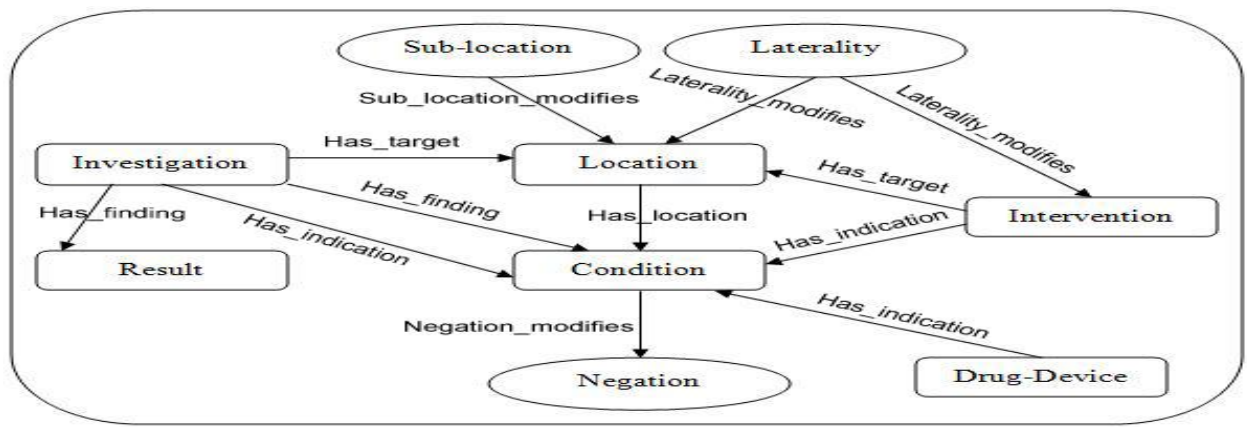

Fig. 2 The relationship schema, showing entities (Rounded rectangles), modifiers (ovals), and relationships (arrows).

## 3.3 Relationship extraction

The GATE NLP toolkit has built our system. GATE NLP toolkit is the tool that allows the applications to be constructed as a pipeline of processing resources [33]. Each resource in this pipeline analyzes the documents, the results of this analysis being available to later resources. The system is shown in Figure 3, and is described below [27]. The pre-processing technique of narratives carried out by using standard GATE modules. The processing resources that used to manipulate the narratives are tokeniser to split narratives into tokens, sentence splitter to split narratives into sentences, part-of-speech (POS) tagged for word tokens, and morphological analyser to find roots for word tokens. POS resource also provides each token with generic POS tag that contains of the first two characters of full POS tag, which called a "generalised" POS tag. After pre-processing technique, guidelines that described in [34] were used assuming that entities extraction is perfect recognition, as given by the entities in the human annotated gold standard described above. The relation extraction depends on the quality of entity extraction.

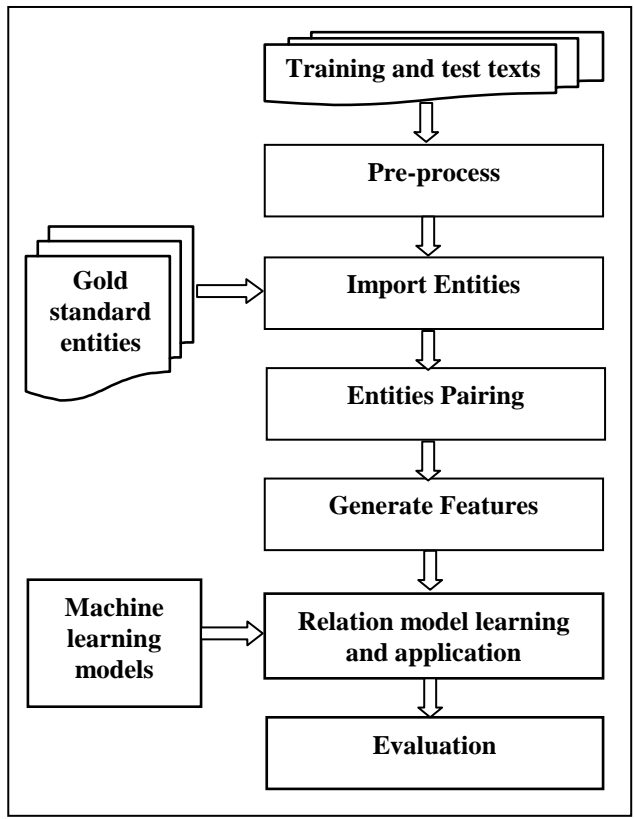

Fig. 3 Relationship extraction system as a GATE pipeline.

### 3.3.1 Classification

The clinical relationship extraction can be manipulated a classification task by assigning relationship type to an entity pair. Pairing the entities that may or may not be the arguments of a relation is called an entity pair. To apply an entity pairing task for each document, all entity pairs that are possible can be created under two constraints. The first constraint, entities pairs must be inside n sentences of each other. For all works in this paper, entities have paired in the same sentence, $n \leq 1$ (crossing 0 or 1 sentence boundaries). The second constraint, entity pairs must be constrained by argument type [35]. For example, there is no relationship between Drug or device entity and a Result entity as specified by the relationship schema. GATE resource developed specifically for extracting relationship from medical texts. This resource also assigns features that characterize lexical and syntactic qualities (described below) of each pair. Entity pairs are compatible with classifier training and test instances. In classifier training, there is two types of results are "class" if an entity pair is compatible with the arguments of a relationship present in the gold standard then there is a class of that relationship type and "class null" if an entity pair is not compatible with the arguments of a relationship. Features of entity pair training instances are used to build a model by the training classifier. In classifier application, unseen text can be used to create entity pairs, under the above constraints. In this classifier, each entity pair assigned class of relationship types or class null [27].

Because of the machine learning algorithms solve a binary-class problem thus to solve a multi-class problem, ML maps this problem to a number of binary classification problems. In multi-class problem the ML plugin implements two common methods are one-against-one and one-against-all. In one-against-one approach each pair of classes require one classifier. In one-against-all approach require a classifier for a binary decision of each pair of the $n$ classes. One-against-all technique is used in our application to solve the multi-class problem.

## 4. Algorithms

There are different machine learning algorithms that can be used for classification to extract relations between entities such as Naïve Bayesian, Decision Tree, K Nearest Neighbour, Perceptron, and Support Vector Machine.

### 4.1 Naive Bayesian

Naive Bayesian classifier is a statistical classifier based on the Bayes' Theorem and the maximum posteriori hypothesis [36]. Let T be a training set of instances. Considering that each data instance to be an n-dimensional vector of attribute values:

$$X = \{x_1, x_2, ..., x_n\} \quad (1)$$

In a Bayesian classifier which assigns each data instance to one of k classes $C_1, C_2, ..., C_k$., a data instance $X$ is assigned to the class for which it has the highest posterior probability conditioned on $X$. This means that, $X$ is assigned to class Ci if and only if

$P(C_i|X) > P(C_j|X)$ for all j such that $1 \leq j \leq n, j \neq i$. (2)

According to Bayes Theorem

$$P(C_i|X) = \frac{P(X|C_i)P(C_i)}{P(X)} \quad (3)$$

Since $P(X)$ is a normalizing factor which is equal for all classes, maximizing the numerator $P(X|C_i)P(C_i)$ is used to do the classification. Values of $P(X|C_i)$ and $P(C_i)$ can be estimated from the data that used to build the classifier.

4.2 Decision Tree

A decision tree is a tree data structure consisting of decision nodes and leaves. A leaf specifies a class value. A decision node specifies a test over one of the attributes, which is called the attribute selected at the node. The C4.5 algorithm constructs the decision tree with a divide and conquers strategy. In C4.5, each node in a tree is associated with a set of cases. Also, cases are assigned weights to take into account unknown attribute values. The C4.5 algorithm uses the concept of information gain or entropy reduction to select the optimal split [37]. Figure 4 shows the pseudo-code of the C4.5 Tree-Construction.

The information gain of an attribute $a$ for a set of cases T is calculated as follow. If $a$ is discrete, and $T_1, ..., T_S$ are the subsets of $T$ consisting of cases with distinct known value for attribute $a$, then:

$$\text{gain} = \text{info}(T) - \sum_{i=1}^{s} \frac{|T_i|}{|T|} \times \text{info}(T_i). \quad (1)$$

Where

$$\text{info}(T) = -\sum_{j=1}^{NClass} \frac{freq(C_j, T)}{|T|} \times \log 2(\frac{freq(C_j, T)}{|T|}) \quad (2)$$

is the entropy function. While having an option to select information gain, by default, however, C4.5 considers the information gain ratio of the splitting $T_1, ..., T_S$, which is the ratio of information gain to its split information:

$$Split(T) = -\sum_{i=1}^{s} \frac{|T_i|}{|T|} \times \log 2(\frac{|T_i|}{|T|}). \quad (3)$$

---

FormTree ($T$)
    (1) ComputeClassFrequency ($T$);
    (2) If OneClass or FewCases
        Return a leaf;
      Create a decision node N;
    (3) ForEach Attribute A
        ComputeGain (A);
    (4) N.test = AttributeWithBestGain;
    (5) If N.test is continuous
        Find Threshold;
    (6) ForEach $T'$ in the splitting of $T$
    (7) If $T'$ is Empty
        Child of N is a leaf
      Else
    (8) Child of N = FormTree ($T'$);
    (9) ComputeErrors of N;
      Return N

Fig. 4 Pseudo-code of C4.5 tree-construction.

4.3 K-Nearest-Neighbor Algorithm (KNN)

K-Nearest-Neighbor classifier is a statistical classifier. When a new sample arrives, $k$-NN finds the K neighbors nearest to the new sample from the training space based on some suitable similarity or distance metric. A common similarity function is based on the Euclidian distance between two data. There are three key elements [38]: a set of labeled objects (e.g., a set of stored records), a distance or similarity metric to compute distance between objects, and the value of $k$, the number of nearest neighbors. To classify an unlabeled object, the distance of this object to the labeled objects is computed, its $k$-nearest neighbors are identified, and the class labels of these nearest neighbors are then used to determine the class label of the object. Figure 5 shows the pseudo-code of the $k$-nearest neighbor classification algorithm. Where $v$ is a class label, $yi$ is the class label for the $i$th nearest neighbors, and $I(\cdot)$ is an indicator function that returns the value 1 if its argument is true and 0 otherwise. A majority vote can be a problem if the nearest neighbors vary widely in their distance and the closer neighbors more reliably indicate the class of the object. Another approach to solve this problem by weighting each object's vote by its distance, where the weight factor:

$$w_i = 1/d(x', x_i)^2 \quad (1)$$

This amounts to replacing the last step of the KNN algorithm with the following: Distance-Weighted Voting:

$$y' = \arg\max_{\upsilon} \sum_{(X_i, y_i) \in D_z} w_i \times I(\upsilon = y_i). \quad (2)$$

---

**Input:** $D$, the set of $k$ training objects and test object $z = (x', y')$

**Process:**

Compute $d(x', x)$, the distance between and every object, $(x, y) \in D$

Select $D_z \subseteq D$, the set of closest training objects to $z$.

**Output:** $y' = \arg\max_{\upsilon} \sum_{(x_i, y_j) \in D_z} I(\upsilon = y_i)$

---

Fig. 5 *K*-Nearest Neighbor classification algorithm.

### 4.4 Perceptron with uneven margin (PAUM)

The advantages of Perceptron with margins are simple, effective, and on-line learning algorithm [39]. The training examples can be checked one by one by predicting their labels by Perceptron. The example is succeeded when the prediction is correct. The example is used to correct the model when the prediction is wrong. The algorithm stops when all training examples are classified by the model correctly. The margin Perceptron has better generalization performance than the standard Perceptron. Figure 6 describes the algorithm of Perceptron with uneven margin.

---

**Require:** A linearly separable training example
$$z = (x, y) \in (X \times \{-1, +1\})^m$$

**Require:** A learning rate $\eta \in R^+$

**Require:** Two margin parameters $\tau_{-1}, \tau_{+1} \in \mathbb{R}^+$

$w_0 = 0; b_0 = 0; t = 0; R = \max_{x_i \in ex} \|X_i\|$

**Repeat**
  **For** $i = 1$ **to** $m$ **do**
    **If** $y_i (\langle w_t, x_i \rangle + b_t) \leq \tau_{y_i}$ **then**
      $w_{t+1} = w_t + \eta y_i x_i$
      $b_{t+1} = b_t + \eta y_i R^2$
      $t \leftarrow t + 1$
    **End if**
  **End for**
**Until** no updates made within the **for** loop
**Return** $(w_t, b_t)$

---

Fig. 6 Algorithm of PAUM $(\tau_{-1}, \tau_{+1})$.

### 4.5 Support vector machine (SVM)

One of the most successful machine learning methods for IE is Support Vector Machine (SVM), which is a general supervised machine learning algorithm. It has achieved state-of-the-art performance on many classification tasks, including named entity recognition [40]. The GATE-SVM system uses a variant of the SVM, the SVM with uneven margins, which has a better generalization performance than the original SVM on imbalanced dataset where the positive examples are much less than the negative ones. Formally, given a training set $Z = ((x_1, y_1), \ldots, (x_m, y_m))$, where $X_i$ is the n-dimensional input vector and $y_i (= +1 \text{ or } -1)$ its label, the SVM with uneven margins is obtained by following the steps in figure 7. In these equations, $\tau$ is the uneven margins parameter which is the ratio of the negative margin to the positive margin in the classifier and is equal to 1 in the original SVM. The goal of the SVM learning is to find the *optimal separating linear hyper-plane* that has the maximum margin linear classifier to both sides.

The SVM problem can be extended to non-linear case using non-linear hyper-plane. Non-linear separation by mapping input data to a high-dimensional space which called kernel function. The new mapping is then linearly separable. Example of kernel function is Polynomial function:

$$K(x, y) = (x^T y + 1)^d \quad (1)$$

---

**Input:** $D$, the set of $m$ training objects $Z = ((x_1, y_1), \ldots, (x_m, y_m))$, and let $y_i \in \{1, -1\}^m$ be the class label of $x_i$.

**Process:** solve the quadratic optimization problem

$$\text{minimise}_{w, b, \xi} \langle W, W \rangle + C \sum_{i=1}^{l} \xi_i \quad (1)$$

subject to
$$\langle W, X_i \rangle + \xi_i + b \geq 1 \quad \text{if } y_i = +1 \quad (2)$$
$$\langle W, X_i \rangle - \xi_i + b \leq -\tau \quad \text{if } y_i = -1 \quad (3)$$
$$\xi_i \geq 0 \quad \text{for } i = 1, \ldots, m \quad (4)$$

**Output:** The decision boundary should classify all points correctly

$$y_i (W^T X_i + b) \geq 1, \quad \forall i$$

---

Fig. 7 Support vector machine algorithm with uneven margin.

## 5. Features for classification

Lexical and syntactic features of tokens and entity pairs that created prior to classification are used to build the classification model. These features are a part of those described in [41] and [42]. These features are split into 14 sets as described in table 2.

TokN features are contained surface string and POS of the tokens that surrounding the entity pairs. This features are provide us with important information about the words surrounding entity pairs to decide if there is relationship between them. GentokN features are generalised tokN which containing morphological root and generalised POS. Str features are contained surface string features include all token features of both entity pairs, their heads, combine of their heads, first, last and other tokens between them, two tokens before the leftmost and after the rightmost entity pairs. POS features are created from POS tags of the entity pairs and the tokens that surrounding them. Root features are created from morphological analyzer of the entity pairs and the tokens that surrounding them. GenPOS features are created from generalised POS tags of the entity pairs and the tokens that surrounding them. Entities were divided into two categories are events and non-events entities. Event entities are Investigation and Intervention entities. Non-event entities are Condition, Location, Drug-device, Result, Negation, Laterality, and Sub-location. Inter features are contained intervening entities which mean types and numbers of entities between entity pairs. Event features are contained whether an entity pairs contain two events, two non-events, or one event and one non-event and if there are any intervening events or non-events between entity pairs. Allgen features are collection of all above features in root and generalised POS forms. Notok features are collection of all above features except for TokN.

Stanford Parser [43] can be applied to parse the corpus to generate a dependency analysis which contains syntactic relations between sentence tokens for the dep and syndist features sets. When the entities exist in the same sentence the dep feature set can be generated from the parse. This feature set consists of the dependency analysis of entity pairs, their heads, and combine of their heads, first, last and other tokens between them, two tokens before the leftmost and after the rightmost entity pairs. For the syndist feature set contains the number of links on the dependency path between the entity pairs and the number of tokens between two entities [28].

## 6. Results and Discussion

Evaluation of the system can be done by using the standard evaluation metrics of Recall and Precision. The terms of true positive (TP), false positive (FP) and false negative (FN) are used to determine Recall and Precision which matches between relations recorded in a system annotated response document and a gold standard key document. If the relation in the response exists in the key with the same arguments then the response relation is a true positive. If the relation in the response dose not exists in the key then the response relation is a false positive. If the relation in the key dose not exists in the response then the key relation is a false negative.

$$R = \frac{TP}{TP+FN} \qquad P = \frac{TP}{TP+FP} \qquad F1 = \frac{2PR}{P+R}$$

Table 2: Feature sets for learning

| Feature set | Description |
|---|---|
| TokN | Surface string and POS of tokens surrounding the arguments, windowed -$N$ to +$N$, $N = 6$ by default. |
| GentokN | Root and generalised POS of tokens surrounding the argument entities, windowed $N$ to +$N$, $N = 6$ by default. |
| Atype | Concatenated semantic type of arguments, in arg1-arg2 order. |
| Dir | Direction: linear text order of the arguments (is arg1 before arg2, or vice versa?). |
| Str | Surface string features based on Zhou et al [29], see text for full description. |
| POS | POS features, as above. |
| Root | Root features, as above. |
| GenPOS | Generalised POS features, as above. |
| Inter | Intervening mentions: numbers and types of intervening entity mentions between arguments. |
| Event | Events: are any of the arguments, or intervening entities, events? |
| Allgen | All above features in root and generalised POS forms, i.e. gen-tok6+atype+dir+root+genpos+inter+event. |
| Notok | All above except tokN features, others in string and POS forms, i.e. atype+dir+str+pos+inter+event |
| Dep | Features based on a syntactic dependency path. |
| Syndist | The distance between the two arguments, along a token path and along a syntactic dependency path. |

Standard ten-fold cross validation methodology is used to split the corpus for evaluation in our experiments. There are scores for each type of relations and for relation overall. P, R and F1 scores are computed for each relation type on each fold and macro-averaging these values for individual relations.

### 6.1 Algorithm type

Multi algorithms are implemented on the training corpus of patient narratives to see which one is much suitable for relation extraction. Many algorithms of supervised machine learning are applied such as Naïve Bayes Weka, C4.5 Weka, KNN Weka, Perceptron algorithm with uneven margin (PAUM), and Support victor machine with uneven margin (SVM), the results of these algorithms are described in table 3.

Table 3: Relation extraction by different algorithms

| Relationship type | Metric (%) | Naive Bayes Weka | C4.5Weka | KNN Weka | PAUM | SVM UM |
|---|---|---|---|---|---|---|
| Has_finding | P | 48.48 | 0.42 | 70.76 | 0.67.5 | 76.66 |
|  | R | 72.11 | 31.16 | 53.11 | 62.04 | 55.85 |
|  | F1 | 52.88 | 29.27 | 52.95 | 58.93 | 57.32 |
| Has_indication | P | 59.85 | 58.67 | 59.77 | 61.71 | 67.17 |
|  | R | 80.39 | 94.79 | 84.76 | 85.59 | 72.85 |
|  | F1 | 67.57 | 71.43 | 68.75 | 70.6 | 68.89 |
| Has_location | P | 67.04 | 66.77 | 69.02 | 71.32 | 76.68 |
|  | R | 94.98 | 95.63 | 91.73 | 92.28 | 85.67 |
|  | F1 | 78.12 | 78.2 | 78.28 | 79.7 | 80.11 |
| Has_target | P | 54.27 | 50.97 | 54.04 | 63.29 | 68.92 |
|  | R | 95.39 | 96.65 | 86.16 | 86.98 | 76.45 |
|  | F1 | 68.71 | 66.36 | 66.01 | 72.88 | 71.59 |
| Laterality_modifies | P | 43.4 | 43.4 | 54.07 | 41.73 | 60 |
|  | R | 58.57 | 58.57 | 68.57 | 58.57 | 51.9 |
|  | F1 | 47.85 | 47.85 | 58.57 | 47.52 | 54.23 |
| Negation_modifies | P | 62.44 | 72.38 | 71.11 | 70 | 80 |
|  | R | 74.16 | 74.16 | 77.5 | 80 | 71.66 |
|  | F1 | 65.73 | 71.84 | 72.72 | 73.75 | 74.66 |
| Sub-location_modifies | P | 77.6 | 90.22 | 92.22 | 90.22 | 100 |
|  | R | 98 | 98 | 98 | 98 | 93 |
|  | F1 | 85.24 | 93.14 | 94.03 | 93.14 | 95.55 |
| Overall | P | 60.95 | 60.81 | 63.33 | 67.34 | 73.99 |
|  | R | 90.51 | 92.18 | 87.63 | 88.59 | 79.67 |
|  | F1 | 72.48 | 73.01 | 73.27 | 76.14 | 76.3 |
| Run Time in seconds |  | 28.563 | 29.999 | 25.148 | 18.736 | 27.487 |

Firstly; Naïve Bayes Weka algorithm is implemented. Different algorithm C4.5 decision tree is applied; overall F1value increases by around 0.5% than the value of NaiveBayesWeka algorithm. KNN Weka algorithm is used with the option '–k 2' to get the best results, there is small increase in the value of overall F1 around by 0.26% than the value of C4.5Weka. Another algorithm PAUM is implemented with the best options "–p 20 –n 5 –optB 0.0". Overall F1 value of PAUM improves the performance than the overall F1value of KNN Weka by around 3%. Finally; SVM with uneven margin algorithm is executed with the options "-c 0.7 -t 1 -d 2 -m 100 -tau 0.8" to get the best results. This means that polynomial kernel is used with degree 2 for quadratic kernel and parameter of uneven margin ($\tau$) is 0.8. There is small change in the overall F1 value of SVM algorithm than overall F1 value of PAUM algorithm around by 0.16%.

From the results in table 3; SVM algorithm with uneven margin is the much suitable machine learning algorithm for relation extraction from medical texts. Figure 8 shows the graph of applying different algorithms for relation extraction from medical texts.

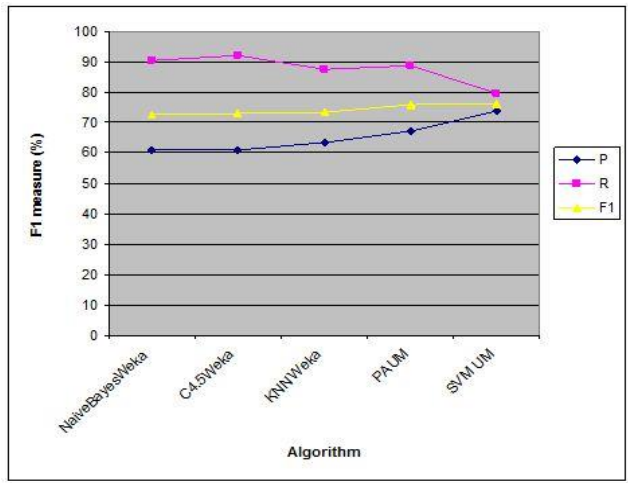

Fig. 8 Performance of different algorithms.

### 6.2 Run time

Different algorithms are implemented; the run time of each algorithm is the most important factor to known which one is much suitable with respect to time to run the application. The run times of each algorithm in seconds described in table 3. Each algorithm is applied on the same features which include the cumulative feature set +event which include different features are *TokN*, *Dir*, *Str*, *POS*, *Inter*, and *Event*.

The C4.5 weka algorithm spends more time to classify the data and extract the relation than other algorithms and the accuracy of the overall F1 measures not perfect very well comparing to other algorithms. The naive bayes weka algorithm needs small time compared to C4.5 weka but the overall F1 value is small than the overall F1 value of C4.5 weka. SVM algorithm is less in time than C4.5 and naïve bayes weka and the F1 measures is greater than these algorithms. This means that SVM is better than C4.5 and Naïve Bayes weka. KNN weka algorithm requires small run time comparing to C4.5, Naïve Bayes, and SVM algorithms. But the accuracy of overall F1 measure is less than SVM and better than C4.5 and Naïve Bayes. The PAUM algorithm considers the faster algorithm than other algorithms and their accuracy F1 measure is better than other algorithms except for SVM algorithm. PAUM algorithm is a faster on small data set than SVM algorithm and there is small difference on the accuracy of the overall F1 measures in between. Figure 9 shows the graph of the run time of each algorithm.

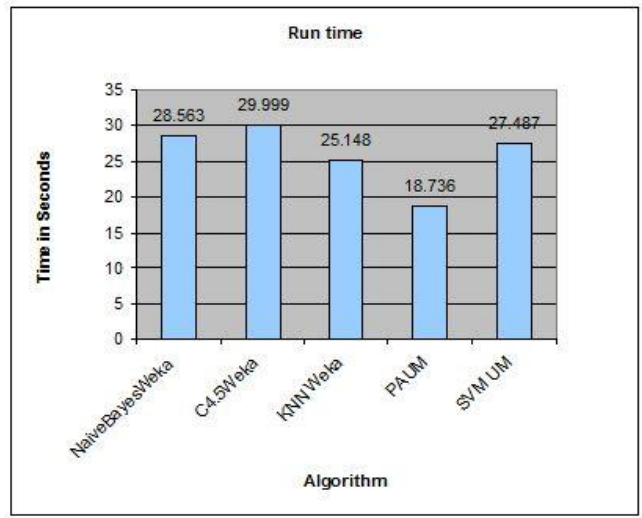

Fig. 9 Run times of different algorithms.

### 6.3 Uneven margin parameter

SVM with uneven margin is the better algorithm in the accuracy than other algorithm but not the faster one. The SVM algorithm is implemented with different uneven parameters to obtain the value of uneven margin ($\tau$) that improves the performance of the system. Then SVM with this uneven margin value is applied with different features to see the effect of adding new feature to the model and also use different corpus size to known their effects on the performance of the system. Table 4 shows SVM with different uneven parameter values. The standard SVM use the uneven margin value 1, this gives bad results than SVM with uneven margin. When the value of uneven margin parameter decreases the results is improved. Figure 10 describes the graph of using different uneven margin parameters.

When the uneven margin value $\tau = 0.8$ is applied, the performance is improved than $\tau = 1$ by around 2.6%. The value of $\tau$ is decreased than this level $\tau = 0.8$ the value of overall F1 measure decreased. Using $\tau = 0.6$, there is small drop on the value of F1 by 0.1%. SVM with $\tau = 0.4$ effects on the performance, this leads to a drop on the value of overall F1 than $\tau = 0.6$ by around 0.36%. When the value of $\tau$ is changed to $\tau = 0.2$, there is drop on the value of overall F1 than $\tau = 0.4$ by around 1.46%. This mean that while SVM is implemented with increasing the value of uneven margin $\tau$, the performance of the system is improved until it is reached to the point that the performance is decreased with increasing the value of $\tau$.

Table 4: SVM use different uneven parameter values

| | Metric (%) | Uneven margin ($\tau$) | | | | |
|---|---|---|---|---|---|---|
| | | 1.0 | 0.8 | 0.6 | 0.4 | 0.2 |
| Overall Relations | P | 76.04 | 73.99 | 69.58 | 65.8 | 61.54 |
| | R | 72.16 | 79.67 | 85.29 | 90.78 | 94.91 |
| | F1 | 73.7 | 76.3 | 76.2 | 75.84 | 74.36 |

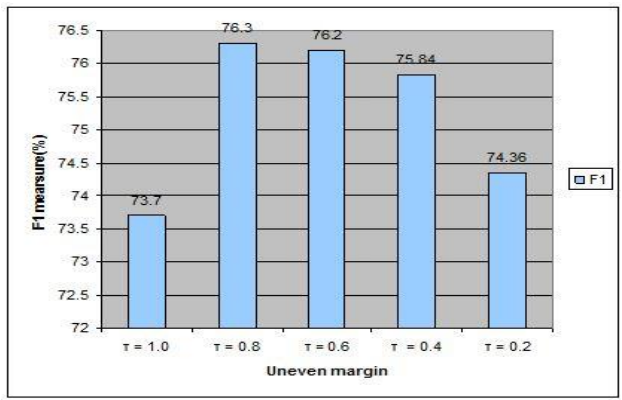

Fig. 10 Performance by different uneven margins.

6.4 Feature selection

The experiments searches for the performance of relation extraction with various feature sets, using the feature sets described in table 2. An additive strategy is used to select the feature. The experiments are divided into two cases, one case of feature sets that do not use syntactic parse information and the other case of feature sets that use syntactic parse information.

6.4.1 Non-syntactic features

Firstly, the experiments used the feature sets that do not use syntactic parse information for relation extraction. Starting with the basic features and then adding new feature set each time to measure the performance of the system. The results are described in table 5.

Starting with Tok6 and Atype features sets, the overall F1 value is 68.29%. Addition of Dir features leads to improve the performance in most metrics, there is improved in the overall F1 value by around 1.35%. Addition of Str features improves the performance in most metrics, there is improved in the overall F1 value by around 0.5%. Addition of the POS features leads to drop the performance in some metrics, overall F1 value drop by around 0.66%. Addition of the Inter features gives more improvements in all metrics, overall F1value increases by around 6.63%. Addition of the Event features gives more improvements in some metrics, overall F1 value increases by around 0.19%.

Generalizing features are used to see their effects on the performance of relation extraction. All Str features, POS features, and TokN features are replaced with their root features, generalized POS features, and generalized TokN features respectively. These results shown in the column Allgen, there is no change in overall F1 value. Notok features are implemented to see if it improves the performance. In this feature TokN features are removed from the full cumulative feature set, corresponding to column +event of table 5. These results are shown in the column Notok, this leads to drop the performance in some metrics, the overall F1value drop by around 0.71%. The graph of using non-syntactic feature sets is shown in figure 11.

6.4.2 Syntactic features

The second part of the feature selection experiments is using features that used syntactic parse information that derived from dependency parse analysis of the texts by using the Stanford parser [43]. The results of +event column in table 5 which corresponding to collection of all non-syntactic feature sets is copied to add in table 6 and then add the Dep features and Syndist features. Addition of the Dep features leads to drop the results that unclear. Addition of the Syndist features leads to a small drop in overall F1 that is unclear. Figure 12 shows the graph of the performance of syntactic feature sets. Addition of the Dep features leads to a drop the performance in some metrics, the overall F1value dropping by around 0.37%. Addition of the Syndist feature set leads to a drop the performance in some metrics, the overall F1 value dropping by 0.38%.

IJCSI International Journal of Computer Science Issues, Vol.10, Issue 1, January 2013
www.IJCSI.org

Table 5: Performance by non-syntactic feature sets

| Relationship type | Metric (%) | Tok6+ Atype | +Dir | +Str | +POS | +Inter | +Event | Allgen | NoTok |
|---|---|---|---|---|---|---|---|---|---|
| Has_finding | P | 5.25 | 55.16 | 66.83 | 45.83 | 61.66 | 76.66 | 76.66 | 74.76 |
|  | R | 43.21 | 48.54 | 45.69 | 29.76 | 42.85 | 55.85 | 55.85 | 58.35 |
|  | F1 | 40.1 | 44.89 | 47.06 | 28.35 | 43.26 | 57.32 | 57.32 | 58.39 |
| Has_indication | P | 64.24 | 62.34 | 63.22 | 62.49 | 64.56 | 67.17 | 67.17 | 66.04 |
|  | R | 68.48 | 69.5 | 70.16 | 71.1 | 69.86 | 72.85 | 72.85 | 71.43 |
|  | F1 | 65.26 | 64.76 | 65.37 | 65.47 | 66.16 | 68.89 | 68.89 | 67.52 |
| Has_location | P | 65.4 | 65.59 | 64.98 | 64.08 | 78.09 | 76.68 | 76.68 | 76.39 |
|  | R | 85.5 | 85.24 | 87.54 | 90.3 | 85.68 | 85.67 | 85.67 | 85.67 |
|  | F1 | 73.38 | 73.39 | 73.91 | 74.4 | 80.82 | 80.11 | 80.11 | 79.94 |
| Has_target | P | 57.04 | 57.47 | 58.64 | 58.18 | 69.1 | 68.92 | 68.92 | 66.7 |
|  | R | 67.52 | 76 | 76.96 | 73.68 | 79.9 | 76.45 | 76.45 | 75.74 |
|  | F1 | 60.08 | 64.31 | 65.52 | 63.77 | 73.46 | 71.59 | 71.59 | 69.99 |
| Laterality_modifies | P | 37.5 | 45.23 | 48.16 | 45 | 60 | 60 | 60 | 60 |
|  | R | 37.14 | 53.57 | 57.14 | 47.14 | 58.57 | 51.9 | 51.9 | 51.9 |
|  | F1 | 36.9 | 48.47 | 50.61 | 45.23 | 59.23 | 54.23 | 54.23 | 54.23 |
| Negation_modifies | P | 70.71 | 70.71 | 70.71 | 70.71 | 75.71 | 80 | 80 | 80 |
|  | R | 76.66 | 76.66 | 70.83 | 70.83 | 71.66 | 71.66 | 71.66 | 71.66 |
|  | F1 | 72.36 | 72.36 | 68.93 | 68.93 | 71.93 | 74.66 | 74.66 | 74.66 |
| Sub-location_modifies | P | 76.54 | 79.6 | 79.6 | 79.6 | 98.33 | 100 | 1.0 | 1.0 |
|  | R | 85 | 93 | 93 | 93 | 93 | 93 | 93 | 93 |
|  | F1 | 78.1 | 83.02 | 83.02 | 83.02 | 94.64 | 95.55 | 95.55 | 95.55 |
| Overall | P | 63.45 | 63.2 | 63.45 | 62.81 | 73.85 | 73.99 | 73.99 | 73.05 |
|  | R | 75.39 | 78.69 | 79.55 | 78.97 | 79.33 | 79.67 | 79.67 | 79.3 |
|  | F1 | 68.29 | 69.64 | 70.14 | 69.48 | 76.11 | 76.3 | 76.3 | 75.59 |

6.5 Size of training corpus

Changing the size of training corpus in the experiments is used to examine their effects on relationship extraction. Two subsets with size 20 and 30 documents is selected from 40 documents; referred to them as C20 and C30, respectively.

The collection feature set of *all* non-syntactic feature sets which represent in +event feature set is used in the experiments to show the effects of training corpus size on the performance, these results are shown in table 7.

Firstly, start the experiments with corpus size 20 documents. Increasing the corpus size to 30 documents leads to improve the performance in most metrics; overall F1 value improves by around 2.11%. Using corpus size 40 documents leads to improve the performance in most metrics; overall F1 value improves by around 2.09%. Increasing the size of the training corpus leads to improve the performance of relation extraction system. Figure 13 shows the effects of changing corpus size in the performance.

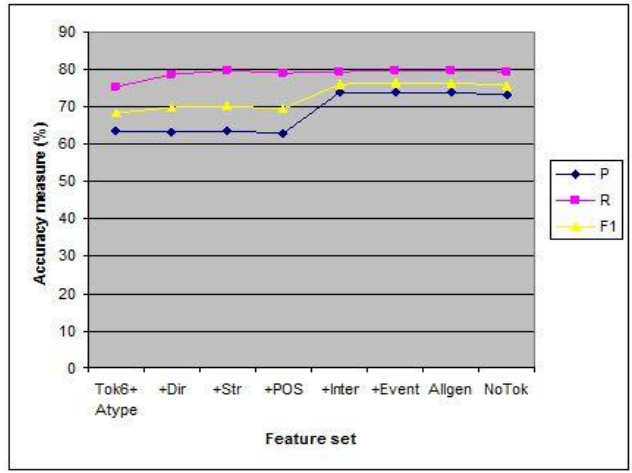

Fig. 11 Graph of non-syntactic feature sets performance.

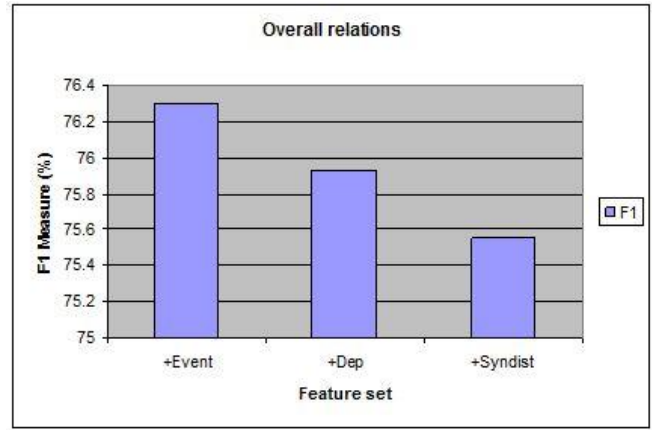

Fig. 12 Graph of syntactic feature sets performance.

Table 6: Performance by Syntactic Feature Sets

| *Relationship type* | *Metric (%)* | *+Event* | *+Dep* | *+Syndist* |
|---|---|---|---|---|
| Has_finding | P | 76.66 | 66.66 | 45 |
|  | R | 55.85 | 44.85 | 17.02 |
|  | F1 | 57.32 | 46.49 | 24.44 |
| Has_indication | P | 67.17 | 66.38 | 65.37 |
|  | R | 72.85 | 71.31 | 68.53 |
|  | F1 | 68.89 | 67.77 | 65.72 |
| Has_location | P | 76.68 | 77.15 | 79.16 |
|  | R | 85.67 | 85.81 | 85.15 |
|  | F1 | 80.11 | 80.52 | 81.12 |
| Has_target | P | 68.92 | 69.23 | 71.56 |
|  | R | 76.45 | 76.84 | 76.31 |
|  | F1 | 71.59 | 72.04 | 73.14 |
| Laterality_modifies | P | 60 | 60 | 50 |
|  | R | 51.9 | 46.9 | 36.90 |
|  | F1 | 54.23 | 50.89 | 40.89 |
| Negation_modifies | P | 80 | 80 | 80 |
|  | R | 71.66 | 71.66 | 71.66 |
|  | F1 | 74.66 | 74.66 | 74.66 |
| Sub-location_modifies | P | 100 | 100 | 100 |
|  | R | 93 | 93 | 93 |
|  | F1 | 95.55 | 95.55 | 95.55 |
| Overall | P | 73.99 | 74 | 74.93 |
|  | R | 79.67 | 78.75 | 76.98 |
|  | F1 | 76.3 | 75.93 | 75.55 |

## 7. Conclusion

From the results, the clinical relationships can be extracted from medical text using different supervised machine learning algorithm. SVM with uneven margin is much suitable algorithm which achieves high accuracy, but it takes more time in the run than Perceptron with uneven margin. Perceptron with uneven margin is very fast algorithm than others as well as the accuracy is relatively near to SVM, there is small change in between. SVM with uneven margin is implemented to show the effects of changing the values of *uneven margin* (τ) parameter, adding the feature sets, and changing the size of the training corpus for relationship extraction. Increasing the value of τ leads to improve the performance to reach the value that has high performance where τ = 0.8 after that point the performance dropped. Adding new feature sets like *non-syntactic features* improves the performance. Adding the *syntactic features* leads to small drop in the performance that unclear. Changing the size of training corpus leads to improve the performance. Our future work on relationship extraction in CLEF includes the integration of a noun and a verb chunk tagger into the feature sets.

Table 7: Performance by corpus size

| Relationship type | Metric (%) | Corpus size | | |
|---|---|---|---|---|
| | | C20 | C30 | C40 |
| Has_finding | Count | 16 | 23 | 36 |
| | P | 46.66 | 10 | 76.66 |
| | R | 21.66 | 3.33 | 55.85 |
| | F1 | 29.04 | 5 | 57.32 |
| Has_indication | Count | 77 | 125 | 180 |
| | P | 58.69 | 65.39 | 67.17 |
| | R | 60.22 | 69.41 | 72.85 |
| | F1 | 58.44 | 65.79 | 68.89 |
| Has_location | Count | 149 | 239 | 363 |
| | P | 75.05 | 79.79 | 76.68 |
| | R | 82.4 | 85.71 | 85.67 |
| | F1 | 77.75 | 82.06 | 80.11 |
| Has_target | Count | 98 | 145 | 180 |
| | P | 69.4 | 69.14 | 68.92 |
| | R | 79.52 | 74.31 | 76.45 |
| | F1 | 72.8 | 70.79 | 71.59 |
| Laterality_modifies | Count | 6 | 9 | 15 |
| | P | 20 | 40 | 0.6 |
| | R | 20 | 30 | 51.9 |
| | F1 | 20 | 33.33 | 54.23 |
| Negation_modifies | Count | 9 | 11 | 20 |
| | P | 40 | 0.5 | 0.8 |
| | R | 40 | 0.45 | 71.66 |
| | F1 | 40 | 46.66 | 74.66 |
| Sub-location_modifies | Count | 11 | 18 | 23 |
| | P | 60 | 80 | 100 |
| | R | 60 | 77.5 | 93 |
| | F1 | 60 | 78.57 | 95.55 |
| Overall | Count | 366 | 570 | 817 |
| | P | 71.1 | 73.36 | 73.99 |
| | R | 74.2 | 75.46 | 79.67 |
| | F1 | 72.1 | 74.21 | 76.3 |

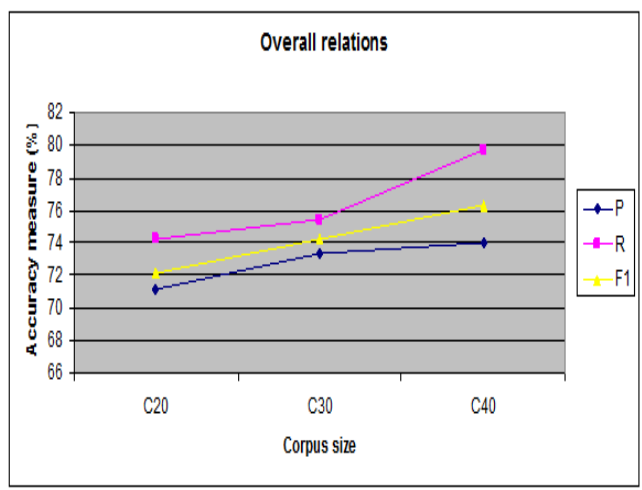

Fig.13. Graph of corpus size performance.

**Wafaa Tawfik Abdel-moneim** is a teaching assistant in information system department in faculty of computers and informatics in zagazig university.

**Mohamed Hashem Abdel-Aziz** is a professor in information system department in faculty of computers and informatics in ain-shames university.

**Mohamed Monier Hassan** a professor assistant in information system department in faculty of computers and informatics in zagazig university.